\documentstyle[preprint,aps]{revtex}
\begin{document}
\draft
\title{Neutrinos and Electromagnetic Gauge Invariance}
\author{F. Pisano\footnote{Supported by FAPESP -- Brazil},    
J.A. Silva-Sobrinho} 
\address{Instituto de F\'\i sica Te\'orica,  
Universidade Estadual Paulista \\  
Rua Pamplona, 145 --  01405-000 -- S\~ao Paulo, SP \\         
Brazil\\}
\author{M.D. Tonasse}
\address{Instituto de F\'\i sica, Universidade do Estado do Rio de Janeiro, \\ 
Rua S\~ao Francisco Xavier, 524 -- 20550-013 -- Rio de Janeiro, RJ \\
Brazil}
\maketitle
\begin{abstract}
It is discussed a recently proposed connection among U(1)$_{\rm em}$      
electromagnetic gauge     
invariance and the nature of the neutrino mass terms in the 
framework of $\mbox {SU(3)}_C\otimes G_W \otimes {\mbox U(1)}_N$, 
$G_W$ = SU(3)$_L$, extensions of the Standard Model. The impossibility of 
that connection, also in the extended case $G_W $ = SU(4)$_L$,   
is demonstrated. 
\end{abstract}
\narrowtext
The neutrinos of the Standard Model are Weyl fermions. The pairs of 
states $(\nu_{lL}, \overline \nu_{lR})$, $l=\{e,\mu,\tau\}$, are the sole 
fermionic neutral non-massive degrees of freedom of the model. However, in a 
large class of extensions of the Standard Model the neutrino can be a massive 
fermion. A mass term for the neutrino can be either Dirac or Majorana, the 
last one being additive violating quantum numbers. If the theory does not 
contain pairs $(\nu_{lR}, \overline\nu_{lL})$ a Dirac mass term is absent, 
but anyway a Majorana mass term is forbidden if the lepton number is a symmetry 
of nature. 
\par
Here we analyze a result contemplated by     
$\ddot {\rm O}$zer~\cite{ozer} which can provide a connection 
between neutrino mass terms and the U(1)$_{\rm em}$ electromagnetic gauge 
invariance. Such a connection comes out in a class of models with gauge 
symmetry
\begin{equation}
G_0 \equiv {\rm SU(3)}_C \otimes G_W \otimes {\rm U(1)}_N,
\label{grupo}
\end{equation}           
where $G_W \equiv {\rm SU(3)}_L, {\rm SU(4)}_L$ are the SU(2)$_L$ weak 
isospin extended groups. Different representation contents are determined by    
embedding the electric charge operator 
\begin{equation}
\frac{Q}{e} = \frac{1}{2}\left (\lambda^L_3 + \xi\lambda^L_8 + 
      \zeta\lambda^L_{15}\right ) + N
\label{carga}
\end{equation}
in the neutral generators $\lambda^L_{3,8,15}$ of $G_W$. The parameters 
$\xi$ and $\zeta$ distinguish different embeddings, fermionic contents, and 
the respective flavordynamics. A salient feature in this class of models  
concerns the anomaly cancellation procedure which is carried out among fermion 
families, color degrees of freedom, and different group transformation 
properties under $G_0$ of the matter field multiplets. As a remarkable result,  
an explanation for the family problem arises.    
\par
The symmetry breaking pattern 
\begin{equation}
G_0\rightarrow {\rm SU(3)}_C \times {\rm SU(2)}_L\times {\rm U(1)_Y} 
\rightarrow {\rm SU(3)}_C\times {\rm U(1)}_{\rm em}
\label{quebra}
\end{equation}      
and masses are achieved by introducing triplets and sextets of scalar    
fields when $G_W = {\rm SU(3)}_L$, or ${\bf 4}$-plets and ${\bf 10}$-plets      
for $G_W = {\rm SU(4)}_L$. 
\par
Let us consider the case in which  
$G_W = {\rm SU(3)}_L$ and therefore $\zeta = 0$ (which we will call 331 
models). In general, three triplets,  
$\phi^a$, and one sextet, $S$, constitute   
the sufficient set of Higgs fields needed to break the original    
gauge symmetry and to generate  masses. Substituting the Gell-Mann matrices 
$$
\lambda^L_3 = {\rm diag}\,\, (+1, -1, 0),  
\quad
\lambda^L_8 = \frac{1}{\sqrt 3}\,{\rm diag}\,\, (+1, +1, -2)
$$          
in the Gell-Mann-Nishijima relation of Eq.~(\ref{carga}) we obtain the 
electric charge of the fields contained in the triplets 
$\phi^a\sim ({\bf 1},{\bf 3}, N_{\phi^a})$, $a = 1,2,3$,   
\begin{equation}
Q(\phi^a) = 
\left (
\begin{array}{c}
\frac{1}{2}(1 + \frac{\xi}{\sqrt 3}) + N_{\phi^a} \\
\frac{1}{2}(-1 + \frac{\xi}{\sqrt 3}) + N_{\phi^a} \\ 
\frac{1}{2}(-\frac{2 \xi}{\sqrt 3}) + N_{\phi^a}
\end{array}
\right )
\label{cargatripl}
\end{equation}
in unities of the proton charge. Performing the Kronecker product 
${\bf 3} \otimes {\bf 3} = {\bf 3}^* \oplus {\bf 6}$ we obtain also the 
electric charges of the fields contained in the symmetric sextet 
$S\sim ({\bf 1},{\bf 6}, N_S)$, 
\begin{equation}
Q(S) = 
\left (
\begin{array}{ccc}
1 + \frac{\xi}{\sqrt 3} + N_S & 
\frac{\xi}{\sqrt 3} + N_S & 
\frac{1}{2} (1 - \frac{\xi}{\sqrt 3}) + N_S \\ 
\frac{\xi}{\sqrt 3} + N_S & 
-1 + \frac{\xi}{\sqrt 3} + N_S & 
-\frac{1}{2}(1 + \frac{\xi}{\sqrt 3}) + N_S \\
\frac{1}{2} (1 - \frac{\xi}{\sqrt 3}) + N_S & 
-\frac{1}{2}(1 + \frac{\xi}{\sqrt 3}) + N_S & 
-\frac{2\xi}{\sqrt 3} + N_S
\end{array}
\right )
\label{cargasex}
\end{equation} 
where 
\begin{equation}
N_S = 2\,N_{\phi^a},  
\label{relation}
\end{equation}
so, we have only three parameters, namely the embedding parameter 
$\xi$ and the U(1)$_N$ charges $N_{\phi^a}$ and $N_S$, in order to know which   
components can develop vacuum expectation values. 
\par
Mass matrices for the electroweak neutral gauge bosons decoupling from 
gluons can be constructed starting with 
the Lagrangian 
\begin{equation}
{\cal L}_{\phi^a,\,S} = 
({\cal D}_\mu \phi^a_i)^\dagger ({\cal D}^\mu \phi^a_i) + 
[({\cal D}_\mu S_{mn})^\dagger ({\cal D}^\mu S_{mn})],  
\label{lagr}
\end{equation}  
with the gauge covariant derivatives 
\begin{equation}
{\cal D}_\mu\phi^a_i = \partial_\mu \phi^a_i - 
ig_L \left ( \vec W_\mu\cdot\vec\lambda^L/2 \right )^j_i \phi^a_j
- i g_N N_{\phi^a_i} B_\mu\phi^a_i 
\label{cov1}
\end{equation}
\begin{equation}
{\cal D}_\mu S_{mn} = \partial_\mu S_{mn} 
-ig_L\left [(\vec W_\mu\cdot \vec\lambda^L/2)^k_m S_{kn} + 
(\vec W_\mu\cdot \vec\lambda^L/2)^k_n S_{km} \right ] 
- i g_N N_S B_\mu S_{mn}
\label{cov2}
\end{equation}
where $\{\vec W_\mu\}$ and $B_\mu$ are the octet and a singlet of gauge bosons 
associated to SU(3)$_L$ and U(1)$_N$ groups and we have denoted the respective 
gauge coupling constants by $g_L$ and $g_N$. The diagonalization of the 
neutral gauge boson mass matrix includes the photon as a linear combination 
of the $W^3_\mu$, $W^8_\mu$ and $B_\mu$ states. If we want to preserve the 
U(1)$_{\rm em}$ local invariance, the inclusion of a mass term for the photon 
$$
\frac{1}{2}M^2_\gamma A_\mu A^\mu
$$   
in the QED Lagrangian is not allowed since 
$$ 
A_\mu A^\mu\rightarrow [A_\mu - \partial_\mu\alpha (x)] 
[A^\mu - \partial^\mu\alpha (x)] \neq A_\mu A^\mu,   
$$
then we are restricted to a singular mass matrix for the electroweak gauge 
bosons. Hence, we demand that the   
contributions to the photon mass in the covariant derivatives coming from any 
scalar field vanish.         
According to Ref. \cite{ozer} this requirement leads to the following 
conditions: 
\begin{equation}
{\rm if}\quad \langle S_{11} \rangle \neq 0 \quad 
{\rm and}\quad N_S\neq 0, \quad{\rm then}\quad 
\langle S_{22} \rangle = 0 \quad{\rm and}
\quad 1 + \frac{\xi}{\sqrt 3} + N_S = 0,
\label{ozer13}
\end{equation}
\begin{equation}
{\rm if}\quad \langle S_{22} \rangle \neq 0 \quad 
{\rm and}\quad N_S\neq 0, \quad{\rm then}\quad 
\langle S_{11} \rangle = 0 \quad{\rm and}
\quad - 1 + \frac{\xi}{\sqrt 3} + N_S = 0,
\label{ozer14}
\end{equation}
\begin{equation}
{\rm if} \quad N_S = 0, \quad {\rm then} \quad 
\langle S_{11} \rangle = \langle S_{22} \rangle = 0.
\label{ozer15}
\end{equation}
In a $G_0$ model with $G_W = \mbox{SU(3)}_L$ the most general 
gauge boson mass matrix is   
\begin{equation}
\frac{1}{2}M^2 = \frac{g_L^2}{4} 
\left (
\begin{array}{ccc}
m_{11} & m_{12} & m_{13} \\
m_{12} & m_{22} & m_{23} \\
m_{13} & m_{23} & m_{33}
\end{array}
\right ) 
\label{matriz3x3}
\end{equation}
where
\begin{eqnarray}
m_{11} & = & \langle\phi^a_i\rangle^2 + \langle\phi^a_j\rangle^2 + 2\left(2
\langle S_{11}\rangle^2 + \langle S_{13}\rangle^2 + 2\langle S_{22}\rangle^2 
+ \langle S_{23}\rangle^2\right) \\ 
m_{22} & = & \frac{1}{3}\left[\langle\phi^a_i\rangle^2 + \langle\phi^a_j
\rangle^2 + 4\langle\phi^a_k\rangle^2\right. \cr
& + &  
\left. 2\left(2\langle S_{11}\rangle^2 + 2\langle S_{22}\rangle^2 + 
8\langle S_{33}\rangle^2 + 4\langle S_{12}\rangle^2 + \langle S_{13}\rangle^2 
+ \langle S_{23}\rangle^2\right)\right ] \\
m_{33} & = & 4t^2\left[N^2_{\phi^a}\left(\langle\phi^a_i\rangle^2 + 
\langle\phi^a_j\rangle^2 + \langle\phi^a_k\rangle^2\right)\right. \cr  
& + & 
\left. N^2_S\left(\langle S_{11}\rangle^2 + 2\langle S_{12}\rangle^2 + 
\langle S_{22}\rangle^2 + 2\langle S_{23}\rangle^2 + 
\langle S_{33}\rangle^2\right)\right] \\ m_{12} & = & \frac{1}{\sqrt 3}
\left[\langle\phi^a_i\rangle^2 -\langle\phi^a_j\rangle^2 + 
2\left(2\langle S_{11}\rangle^2 - 2\langle S_{22}\rangle^2 - 
\langle S_{13}\rangle^2 + \langle S_{23}\rangle^2\right)\right]
\label{ninguem1} \\
m_{13} & = & 2t\left[N_{\phi^a}\left(\langle\phi^a_i\rangle^2 - 
\langle\phi^a_j\rangle^2\right) + 2N_S\left(\langle S_{11}\rangle^2 - 
\langle S_{22}\rangle^2 + \langle S_{13}\rangle^2 - \langle S_{23}
\rangle^2\right)\right] \\
m_{23} & = & \frac{2t}{\sqrt 3}\left[N_{\phi^a}\left(\langle\phi^a_i
\rangle^2 + \langle\phi^a_j\rangle^2 - 2\langle\phi^a_k\rangle^2\right)
\right. \cr  
& + &  
\left. 2N_S\left(\langle S_{11}\rangle^2 + \langle S_{22}\rangle^2 - 2
\langle S_{33}\rangle^2 + 2\langle S_{12}\rangle^2 - \langle S_{13}\rangle^2 
- \langle S_{23}\rangle^2\right)\right]  
\label{elementos}
\end{eqnarray}
with $t\equiv g_N/g_L$ and $i,j,k$ are SU(3) indices.     
\par
Let us consider the model with the embedding parameter 
$\xi = -\sqrt 3$~\cite{pp}. Leptons transform under 331 as    
\begin{equation}
\Psi_{aL} = 
\left ( 
\nu_{l_a}, \,\,\,  
l_a, \,\,\, 
l^c_a
\right )^T_L \sim ({\bf 1},{\bf 3},0)
\label{pplep}
\end{equation} 
where $a = 1,2,3$ is the family index and $l^c_a$ is the charge      
conjugate field corresponding to $l_a$. The scalar sector is the 
set of three SU(3)$_L$ triplets     
$\eta = (\eta^0,\,\,\eta^-_1,\,\,\eta^+_2)^T \sim ({\bf 1},{\bf 3},0)$, 
$\rho = (\rho^+,\,\,\rho^0,\,\,\rho^{++})^T \sim ({\bf 1},{\bf 3},+1)$,  
$\chi = (\chi^-,\,\,\chi^{--},\,\,\chi^0)^T \sim ({\bf 1},{\bf 3},-1)$ 
and the sextet 
\begin{equation}
S = 
\left (
\begin{array}{ccc}
\sigma^0_1 & h^-_2 & h^+_1 \\
h^-_2 & H^{--}_1 & \sigma^0_2 \\
h^+_1 & \sigma^0_2 & H^{++}_2
\end{array}
\right ) 
\sim ({\bf 1},{\bf 6},0)
\label{sex}
\end{equation}
whose neutral components can develop a vacuum expectation value. A nonvanishing 
$\langle\sigma^0_1\rangle$ gives mass to neutrinos. The gauge boson mass       
matrix elements are    
\begin{eqnarray}
m_{11} & = & \langle\eta\rangle^2 + 
\langle\rho\rangle^2 + 
4\langle{\sigma_1}\rangle^2 + 2\langle{\sigma_2}\rangle^2 
 \\
m_{22} & = & \frac{1}{3}\left (\langle{\eta}\rangle^2 + 
\langle{\rho}\rangle^2 + 4\langle{\chi}\rangle^2 + 
4\langle{\sigma_1}\rangle^2 + 2\langle{\sigma_2}\rangle^2 \right ) 
\\
m_{33} & = & 4t^2\left(\langle\rho\rangle^2 + \langle\chi\rangle^2\right) \\
m_{12} & = & \frac{1}{\sqrt 3}\left(\langle\eta\rangle^2 - 
\langle\rho\rangle^2 + 4\langle{\sigma_1}\rangle^2 
+ 2\langle{\sigma_2}\rangle^2\right) \\
m_{13} & = & -2t\langle\rho\rangle^2 \\
m_{23} & = & \frac{2}{\sqrt 3}t\left(\langle\rho\rangle^2 + 
2\langle\chi\rangle^2\right) 
\end{eqnarray}
This is a singular matrix even if $\langle S_{11} \rangle = 
\langle{\sigma^0_1}\rangle \neq 0$.   
\par
The scalar potential contains terms as 
\begin{equation}
V(\eta,\rho,\chi,S) = ...+ f_1\eta^\dagger S\eta^* + f_2{\rm det} S + 
f_3\epsilon_{ijk}(S\eta^*)_i\rho_j\chi_k + 
f_4\epsilon_{ijk}\epsilon_{lmn}S_{il}S_{jm}\rho_k\chi_n.
\label{daniel1}
\end{equation}
If we wish to avoid tree level neutrino masses,  
$\langle\sigma^0_1\rangle = 0$, we have a fine-tuning among the 
$f_{1,...,4}$ coupling constants.   
In order to avoid fine tuning between the coupling constants of 
the most general Higgs potential $V\left(\eta , \rho , \chi , S\right)$ and 
maintain lepton and baryon number conservation the following discrete 
symmetries must be introduced if $\langle \sigma^0_1 \rangle = 0$~\cite{daniel}:
\begin{eqnarray}
Q_{1L} & \rightarrow & -Q_{1L}, \quad \eta \rightarrow -\eta, \nonumber \\
Q_{jL} & \rightarrow & -iQ_{jL}, \quad \rho, \,\, \chi \rightarrow i\rho, \,\, 
\chi \nonumber \\
\Psi_{lL} & \rightarrow & i\Psi_{lL}, \quad S \rightarrow -S, 
\label{sym}\\
u_{jR} & \rightarrow & u_{jR}, \quad J_{1R} \rightarrow iJ_{1R}, \nonumber \\
d_{jR} & \rightarrow & id_{jR}, \quad J_{2,3R} \rightarrow J_{2,3R}, 
\nonumber
\end{eqnarray}
where $Q_{1L}$, $Q_{jL}$ ($j = 2, 3$) are SU(3)$_L$ triplets of quarks, and 
$J_{1,2,3R}$ are the corresponding right-handed exotic quark 
singlets~\cite{pp}. The leptonic Yukawa couplings involving the Higgs sextet 
$S\sim ({\bf 1},{\bf 6},0)$ allowed by the gauge invariance have     
the general form     
\begin{equation}
{\cal L}_{l,S} = -\frac{1}{2}\sum_{l,m}G_{lm}\overline{\Psi^c}_{il}
\Psi_{jm}S_{ij} + 
\mbox{H.c.}     
\label{yuk1}
\end{equation}
where $i$, $j$ denote SU(3) indices, $l,m = e,\mu,\tau$, and 
$G_{lm}=G_{ml}$. In the $\xi = -\sqrt 3$ model these couplings 
for the neutrinos are 
\begin{equation}
{\cal L}_{\nu,S} = -\frac{1}{2}\sum_{l,m}G_{lm}
\left (\bar{\nu^c}_{lR}\nu_{mL}S_{11} + \mbox{H.c.} \right ). 
\label{yuk2}
\end{equation}
Then Majorana mass terms are not allowed if      
$\langle S_{11}\rangle = \langle\sigma^0_1\rangle = 0$.    
Notwithstanding such condition is entirely independent and does not affect 
the mass spectrum pattern of the gauge sector.   
\par 
In the $\xi = 1/\sqrt 3$ model~\cite{mpp}, with a symmetric sextet 
$S\sim ({\bf 1},{\bf 6},+2/3)$, the Yukawa couplings are    
\begin{equation}
{\cal L}_{\nu,S} = -\frac{1}{2}\sum_{l,m}G_{lm}
\left (\bar{\nu}_{lR}\nu^c_{mL}S_{11} + \bar{\nu}_{lR}\nu_{mL}S_{12}       
+ \bar{\nu^c}_{lR}\nu^c_{mL}S_{12} + \bar{\nu^c}_{lR}\nu_{mL}S_{22}         
+ \mbox{H.c.}\right ).
\label{yuk3}
\end{equation}
The sufficient scalar sector of the $\xi=1/\sqrt{3}$ model contains the set of  
three triplets $\eta = (\eta^0, \eta^-_1, \eta^-_2)^T 
\sim ({\bf 1},{\bf 3},-2/3)$, $\rho = (\rho^+, \rho^0_1,\rho^0_2)^T 
\sim({\bf 1},{\bf 3},1/3)$, and   
$\chi = (\chi^+, \chi^0_1,\chi^0_2)^T$ 
with the same transformation properties as the $\rho$ multiplet.  
The corresponding gauge boson mass matrix given in  
Eq.~(\ref{matriz3x3}) including the sextet 
\begin{equation}
S = \left (
\begin{array}{ccc}
S^0_1 & S^0_2 & S^-_1 \\
S^0_2 & S^0_3 & S^-_2 \\
S^-_1 & S^-_2 & S^{--}
\end{array}
\right ) 
\sim 
({\bf 1},{\bf 6},+2/3)
\nonumber
\end{equation}
has the elements 
\begin{eqnarray}
m_{11} & = & \langle\eta\rangle^2 + \langle\rho_1\rangle^2 + 
              \langle\chi_1\rangle^2 + 4\langle S_1\rangle^2 + 
             \langle S_3\rangle^2 \\
m_{22} & = & \frac{1}{3}[m_{11} + 4 (\langle\rho_2\rangle^2 + 
              \langle\chi_2\rangle^2 + 2\langle S_2\rangle^2) + 
             3\langle S_3\rangle^2] \\
m_{33} & = & \frac{4}{9}t^2[m_{11} + 3\langle\eta\rangle^2   
              + \langle\rho_2\rangle^2 + 8 \langle S_2\rangle^2 + 
              \langle\chi_2\rangle^2 + 3\langle S_3\rangle^2] \\
m_{12} & = & \frac{1}{\sqrt 3}(\langle\eta\rangle^2 - \langle\rho_1\rangle^2 
             - \langle\chi_1\rangle^2 + \langle S_1\rangle^2 
             - \langle S_3\rangle^2) \\
m_{13} & = & \frac{4}{3}t(-\langle\eta\rangle^2 - 
             \frac{1}{2}\langle\rho_1\rangle^2 - 
             \frac{1}{2}\langle\chi_1\rangle^2 + 2\langle S_1\rangle^2 
             -2\langle S_3\rangle^2) \\
m_{23} & = & \frac{4}{3\sqrt 3}t(-\langle\eta\rangle^2 + \frac{1}{2} 
             \langle\rho_1\rangle^2 - \langle\rho_2\rangle^2 - 
             \langle\chi_2\rangle^2 + 2\langle S_1\rangle^2 + 
             4\langle S_2\rangle^2 + 2\langle S_3\rangle^2). 
\label{matrizmpp}
\end{eqnarray} 
This matrix is singular only if hold the conditions $\langle S_{11}\rangle = 
\langle S_{12}\rangle = \langle S_{22}\rangle = \langle\chi_1\rangle = 0$, so 
neutrinos are massless at tree level. Another possible Yukawa couplings in 
this model are
\begin{equation}
{\cal L}_{l,\eta} = -\frac{1}{2}\sum_{l,m}\epsilon^{ijk}h_{lm}
\Psi_{il}C^{-1}\Psi_{jm}\eta^*_k
\end{equation}   
with $C$ being the charge conjugation operator and $h_{lm}=-h_{ml}$ which 
implies an antisymmetric $3\times 3$ mass matrix 
for the neutrinos. Hence, there are one massless and two mass 
degenerate neutrino states, at least at tree level.  
\par
For the $\xi = -1/\sqrt 3$, $\zeta = -2\sqrt 6/3$ model~\cite{su4} with 
$G_W = \mbox{SU(4)}_L$ the neutrino Yukawa couplings are  
\begin{equation}
{\cal L}_{\nu,H} = -\frac{1}{2}\sum_{l,m}G_{lm}
\left (\bar{\nu^c}_{lR}\nu_{mL}H_{11} + \bar\nu_{lR}\nu_{mL}H_{13}        
+ \bar{\nu^c}_{lR}\nu^c_{mL}H_{13} + \bar\nu_{lR}\nu^c_{mL}H_{33}       
+ \mbox{H.c.}\right )
\label{yuk4}
\end{equation}
where the $H_{ij}$, $i,j = 1,...,4$ are the elements of the SU(4) symmetric  
{\bf 10}-plet 
\begin{equation}
H = \left (
\begin{array}{cccc}
H^0_1 & H^+_1 & H^0_2 & H^-_2 \\
H^+_1 & H^{++}_1 & H^+_3 & H^0_3 \\
H^0_2 & H^+_3 & H^0_4 & H^-_4 \\
H^-_2 & H^0_3 & H^-_4 & H^{--}_2 
\end{array}
\right ) \sim ({\bf 1},{\bf 10},0).
\label{decupleto}
\end{equation}
The vacuum structure $\langle H^0_3 \rangle \neq 0$, 
$\langle H^0_{1,2,4}\rangle = 0$ is sufficient for giving a finite mass to the  
charged leptons but neutrinos remain massless at tree level. 
\par
The mass matrix elements of the gauge bosons in the model with 
$G_W = \mbox{SU(4)}$, up to a factor $g^2_{SU(4)}/4$, are   
\begin{eqnarray}
m_{11} & = & \langle\eta_1\rangle^2 + \langle\eta^\prime_1\rangle^2 +
             \langle\rho\rangle^2 + 2\left(2\langle H_1\rangle^2 +
             \langle H_2\rangle^2 + \langle H_3\rangle^2\right) \\
m_{22} & = & \frac{1}{3}\left[m_{11} + 4\left(\langle\eta_2\rangle^2
             + \langle \eta^\prime_2\rangle^2 + 
               4\langle H_4\rangle^2\right)\right] \\
m_{33} & = & \frac{1}{6}\left[m_{11} + \langle\eta_2\rangle^2 + 
             \langle\eta^\prime_2\rangle^2 + 9\langle\chi\rangle^2 +
             6\left(\langle H_2\rangle^2 + \langle H_3\rangle^2 \right ) +
             4\langle H_4\rangle^2\right] \\
m_{44} & = & 4t^2 \left(\langle\rho\rangle^2 + \langle\chi\rangle^2\right) \\
m_{12} & = & \frac{1}{\sqrt 3}\left[\langle\eta_1\rangle^2 + 
             \langle\eta^\prime_1\rangle^2 - \langle\rho\rangle^2 +  
             2\left(2\langle H_1\rangle^2 -  \langle H_2\rangle^2  
             - \langle H_3\rangle^2 \right)\right] \\
m_{13} & = & \frac{1}{\sqrt 6}\left[\langle\eta_1\rangle^2 +
             \langle\eta^\prime_1\rangle^2 - \langle\rho\rangle^2 +
             4\left(\langle H_1\rangle^2 + \langle H_2\rangle^2 +
             \langle H_3\rangle^2\right)\right] \\
m_{14} & = & -2t\langle\rho\rangle^2 \\
m_{23} & = & \frac{1}{3\sqrt 2}\left[\langle\eta_1\rangle^2 +  
             \langle\eta^\prime_1\rangle^2 - 2\left(\langle\eta_2\rangle^2
             + \langle\eta^\prime_2\rangle^2\right) + \langle\rho\rangle^2
             + 4\left(\langle H_1\rangle^2 - \langle H_2\rangle^2 - 
             \langle H_3\rangle^2 - 2\langle H_4\rangle^2\right)\right] \\
m_{24} & = & \frac{2}{\sqrt 3}t\langle\rho\rangle^2 \\
m_{34} & = & \frac{2}{\sqrt 6}t\left(\langle\rho\rangle^2 +
             3\langle\chi\rangle^2\right) 
\label{su4elem}
\end{eqnarray}
where $\eta\sim ({\bf 1},{\bf 4}, 0)$, $\eta^\prime \sim ({\bf 1},{\bf 4}, 0)$, 
$\rho \sim ({\bf 1},{\bf 4}, +1)$, and $\chi \sim ({\bf 1},{\bf 4}, -1)$. 
This is the most general $G_W = \mbox{SU(4)}_L$ electroweak    
gauge boson mass matrix.    
As it can be checked, it is a singular matrix for any 
$\langle H^0_{ij} \rangle \neq 0$.    
None of the $H$ matrix elements imposes any constraint on the neutrino Yukawa  
couplings of Eq.~(\ref{yuk4}). 
\par
We outline our results. The constraint given in Eq. (\ref{ozer15}) is not 
realized in the $G_0$ models. This is true also in the 331 model of 
Ref.~\cite{frampton} which is equivalent to the model of Ref.~\cite{pp}. 
In fact these two representations are identical since we can map from 
one representation to another by a unitary transformation~\cite{ng}. 
It can be checked that condition in Eq. (\ref{ozer14}) does not hold 
in the $\xi = 1/\sqrt 3$ model since for the nonvanishing vacuum 
expectation value of any neutral component   
of the sextet with $N_S = 2/3$ the electromagnetic gauge 
invariance is not   preserved. As we have shown, also in the 
$G_W = \mbox{SU(4)}_L$ extensions, there is no connection among 
neutrino masses and electromagnetic gauge invariance. However, a 
remarkable fact is that a possible constraint on 
neutrino masses in the $\xi = -\sqrt 3$ model can be arisen from the 
scalar potential through the set of discrete symmetries of 
Eqs. (\ref{sym}) in order to avoid a fine tuning among the scalar 
potential coupling constants which is consistent for 
$\langle\sigma_1^0\rangle = 0$. For concluding we wish to point out     
that in a large class of gauge models which contain the right-handed neutrino  
the electric charge quantization can be obtained only if the neutrino   
is a Majorana particle~\cite{bm}. This fact is true even for the Standard 
Model enlarged for containing a right-handed neutrino.   
Thus in the $G_0$ models, since the conditions of Eq. (\ref{ozer15}) are 
not true, there is not contradiction with this fact.



\end{document}